\documentclass[12pt]{article}

\def\lsi{\raise0.3ex\hbox{$<$\kern-0.75em\raise-1.1ex\hbox{$\sim$}}}
\def\gsi{\raise0.3ex\hbox{$>$\kern-0.75em\raise-1.1ex\hbox{$\sim$}}}

\newcommand{\be}{\begin{equation}}
\newcommand{\ee}{\end{equation}}
\newcommand{\nn}{\nonumber}
\newcommand{\bea}{\begin{eqnarray}}
\newcommand{\eea}{\end{eqnarray}}
\newcommand{\eps}{\epsilon}

\newcommand{\R}{{\kern+.25em\sf{R}\kern-.78em\sf{I} \kern+.78em\kern-.25em}}
\newcommand{\N}{{\kern+.25em\sf{N}\kern-.78em\sf{I} \kern+.78em\kern-.25em}}
\newcommand{\C}{{\kern+.25em\sf{C}\kern-.50em\sf{I} \kern+.50em\kern-.25em}}

\def\gmu{\gamma_{\mu}}
\def\gnu{\gamma_{\nu}}

\def\Tr{\mbox{Tr}}
\def\tr{\mbox{tr}}
\def\e{\epsilon}
\def\g5{\gamma_{2n+1}}
\def\A{{\cal A}}

\def\O{{\cal O}}

\def\C{{\cal C}}

\def\N{{\cal N}}

\def\lb{\lbrack}
\def\rb{\rbrack}

\makeatletter
\@addtoreset{equation}{section}
\makeatother

\begin{document}

\begin{flushright}

INLO-PUB-10/03 \\
HU-EP-03/38 \\
SFB-CPP-03-14 \\

\end{flushright}

\vspace*{5mm}

\begin{center}
{\LARGE\bf Axial Anomaly and Index of} \\

\vspace*{6mm}

{\LARGE\bf the Overlap Hypercube Operator} 

\vspace*{1cm}

David H. Adams$^{\rm \, a}$ \ and \ Wolfgang Bietenholz$^{\rm \, b}$

\vspace*{8mm}

$^{\rm a}$ Instituut-Lorentz for Theoretical Physics,
Leiden University, \\
Niels Bohrweg 2, NL-2333 CA Leiden, The Netherlands\\

\vspace*{3mm}

$^{\rm b}$ Institut f\"{u}r Physik, Humboldt Universit\"{a}t zu Berlin \\
Newtonstr.\ 15, D-12489 Berlin, Germany \\

\end{center}

\vspace*{5mm}

The overlap hypercube fermion is constructed by inserting
a lattice fermion with hypercubic couplings into the overlap 
formula. One obtains an exact Ginsparg-Wilson fermion, which
is more complicated than the standard overlap fermion, but which
has improved practical properties and is of current interest
for use in numerical simulations.
Here we deal with conceptual aspects of the overlap hypercube 
Dirac operator. Specifically, we evaluate the axial anomaly and 
the index, demonstrating that the correct classical continuum limit 
is recovered. Our derivation is non-perturbative and therefore valid 
in all topological sectors. At the non-perturbative level this result
had previously only been shown for the standard overlap Dirac operator 
with Wilson kernel. The new techniques which we develop to accomplish
this are of a general nature and have the potential to
be extended to overlap Dirac operators with even more general kernels.

\vspace*{4mm}

\noindent
PACS: 11.15.Ha, 11.30.Rd \\
Keywords: Lattice fermions, Ginsparg-Wilson relation,
chiral anomaly, Index Theorem \\
hep-lat/0307022

\newpage

\section{Introduction}

In gauge theories with fermions, the index of a Dirac operator 
plays an important r\^{o}le. It is given as the difference of the
number of left- and right-handed zero modes, and
due to the Atiyah-Singer Index Theorem \cite{AS} it can be identified 
with the topological charge of the gauge field. 
Therefore the same quantity also provides the integrated axial anomaly.

On the lattice it is a non-trivial question if these quantities
can be recovered. In particular it is not clear a priori
if we obtain the correct axial
anomaly in the classical continuum limit. The traditional lattice
formulation by means of the Wilson Dirac operator $D_{W}$
does not allow for the Index Theorem to be adapted: 
lattice gauge configurations do not have natural
topological sectors, and the Wilson fermion does not have exact zero
modes. Nevertheless the axial anomaly can be reproduced \cite{Wil};
a necessary condition for this property is the absence of species
doublers.

The situation is different for overlap fermions \cite{HN}. 
They have good chiral properties according to their origin
from the overlap formalism \cite{NN,RDS}, which are reflected by the
fact that their lattice Dirac operator $D_{ov}$ obeys the 
Ginsparg-Wilson relation (GWR) \cite{GW,Has}. 
In an even dimension $2n$ it reads
\footnote{We refer to a Euclidean lattice of spacing $a$, and the
fermions belong to some unitary representation of the (unspecified)
gauge group.}
\be  \label{GWR}
D_{ov} \gamma_{2n+1} +\gamma_{2n+1} D_{ov} =
\frac{a}{m} D_{ov} \gamma_{2n+1} D_{ov} \ .
\ee
Here $m$ is a parameter which controls topological properties
and the number of fermion species described by the corresponding
lattice fermion action.
The GWR turns into the standard
condition for chiral symmetry in the continuum limit.
Moreover, even on the lattice an exact chiral symmetry exists \cite{ML}.
It is lattice modified by a local term of $O(a)$.
Explicitly, the variation of the spinor fields under
the lattice modified chiral transformation
can be written in the form \cite{ML}
\be  \label{trafo}
\delta \bar \psi = \bar \psi \Big( 1-\frac{a}{2m} D_{ov} \Big) \gamma_{2n+1}\ , 
\qquad \delta \psi = \gamma_{2n+1} \Big( 1-\frac{a}{2m} D_{ov} \Big) \psi \ ,
\ee
which leaves the lattice action invariant.

The Ginsparg-Wilson relation excludes
additive mass renormalization. The index is well defined,
since the exact zero modes have a definite chirality \cite{Has}.
\footnote{In the case of the overlap Dirac operator the index 
coincides with the overlap topological charge \cite{NN}.
For the fixed point fermion, which also solves the GWR,
it coincides with the classically perfect topological charge 
\cite{Has}.}
The question arises if the correct axial anomaly
is reproduced in the continuum limit.
In fact, the existence of some anomaly is obvious since the
fermionic measure is not invariant under the transformation
(\ref{trafo}). The resulting anomaly takes the form
\be
{\cal A} (x) = 2i q(x) \ ,
\ee
where $q(x)$ is the topological charge density,
\be  \label{density}
q(x) = -\frac{1}{2m} \tr \Big( \gamma_{2n+1}D_{ov} \Big)(x,x) \ .
\ee
Generally we define the density $\O (x,y)$ of a lattice operator
$\O$ by 
\be
\O\psi(x) = a^{2n} \sum_y \O(x,y)\psi(y) \ .
\ee 
Then the operator trace can be expressed as
\be
\Tr \, \O = a^{2n} \sum_x \tr \, \O (x,x) \ ,
\ee
where $\tr$ sums over spinor and gauge indices.
In particular the index is given by
\be  \label{index}
{\rm index} (D_{ov}) = - \frac{1}{2 m}
\Tr \, (\gamma_{2n+1} D_{ov}) = a^{2n} \sum_{x} q(x) \ .
\ee

The question if one obtains the correct expression for the axial anomaly
has been studied in a number of papers. This issue was first addressed
in Ref.\ \cite{GW}, where a perturbative evaluation of the anomaly was given
for a general Dirac operator satisfying the Ginsparg-Wilson relation.
\footnote{This perturbative evaluation was reconsidered in the context of 
modern developments of the Ginsparg-Wilson relation \cite{CH}.}
Furthermore, a perturbative evaluation of the axial anomaly for 
very general
lattice Dirac operators has been carried out in Refs.\ \cite{RR}.
In all cases the anomaly was found to reduce to the correct continuum
expression if the operator is local and free of species doubling.
However, the use of perturbation theory restricts the validity of these
arguments to the sector of topologically trivial gauge fields.
In fact, the question for which Ginsparg-Wilson fermions
the correct anomaly holds also in topological sectors different
from zero seems to be non-trivial in the light of Ref.\ \cite{TWC},
which presents an example where this is apparently not the case.
So far, the {\em only} Ginsparg-Wilson operator for which the correct 
continuum limits of the axial anomaly and index have been established
{\em non-perturbatively} is the standard overlap Dirac operator, 
which is given by substituting the Wilson-Dirac operator $D=D_W$ 
into the overlap formula
\be  \label{over}
D_{ov} = \frac{m}{a} \Big[ 1 + A / \sqrt{A^{\dagger}
A} \ \Big] \ , \quad A = D - \frac{m}{a} \ .
\ee
The continuum limit of the axial anomaly in this case has been studied 
explicitly in Refs.\ \cite{KYFS,DA(AP),DA(JMP),FNS}.
In particular, a rigorous non-perturbative demonstration that
the anomaly and the index have the correct continuum limit in 
all topological sectors when $m$ is in the physical (doubler-free) region 
was given in \cite{DA(AP),DA(JMP)}. 

Currently there is interest in non-standard overlap Dirac operators 
obtained by inserting more general lattice Dirac operators $D$ into
the overlap formula (\ref{over}). The background and motivation for 
this is discussed further below. In particular, overlap Dirac operators
where the input $D$ is a {\em hypercubic fermion} (HF) operator
have been the focus of attention.\footnote{HF operators are generalizations
of the Wilson-Dirac operator which couple all sites within a lattice 
hypercube. (Recall that the Wilson-Dirac operator couples only nearest
neighbor sites.)} In the light of this development it is pertinent
to show at the non-perturbative level that the axial anomaly and index 
for these non-standard overlap Dirac operators also have the correct 
continuum limit in all topological sectors. As mentioned above,
this has so far only been shown for the standard overlap operator 
where the input $D$ is the Wilson-Dirac operator.
The purpose of the present paper is to establish this result at the
non-perturbative level for non-standard overlap Dirac 
operators of specific current interest, namely those for which
the input $D$ in the overlap formula is a HF operator. 
To do this we follow the rigorous
non-perturbative approach of Refs.\ \cite{DA(AP),DA(JMP)} for the 
standard overlap Dirac case, and take inspiration from Ref.\ \cite{FNS},
where a topological description of the anomaly coefficient
as the degree of a certain map was derived which greatly facilitates its
evaluation. However, the key technical parts of the 
arguments in those papers are specific to the standard overlap case ---
they rely on the explicit form of the Wilson kernel
and do not have a straightforward generalization to more general 
kernels. Therefore we have had to develop new techniques and formulae
to handle the more general HF case. In fact, our techniques are of a 
general nature and have the potential to be used for even more general 
kernels. (The case of overlap Dirac operator with completely general kernel
has further complications though, and is postponed to a later analysis.)

Let us now discuss the background and motivation for considering the 
non-standard overlap operators mentioned above. Their use was suggested 
in Ref.\ \cite{EPJC}. The motivation is to improve other properties of 
the overlap Dirac operator --- beyond chirality ---
which are also of importance for a lattice fermion formulation,
such as the quality of scaling, locality and of approximate
rotation invariance. We emphasize that chiral properties of the overlap
operator $D_{ov}$ continue to hold for {\em any} input lattice Dirac 
operator $D$ (free of species doubling) in (\ref{over}), since $D_{ov}$
satisfies the GWR for any such choice. The basic idea is to construct
a short range, doubler-free lattice Dirac operator $D$ for the input into
the overlap formula which has the following properties: (i) good
scaling and approximate rotational invariance, and (ii) good
chirality in the sense that $D$ is an approximate solution of
the GWR. The property (ii) suggests that the overlap Dirac operator $D_{ov}$
obtained from inserting $D$ into the overlap formula will inherit to a large
degree the properties (i) of $D$, and will furthermore have good locality
properties. To see this, note 
that if $D$ is an exact solution of the GWR then the overlap formula just
gives $D$ back again: $D_{ov}=D$ \cite{EPJC}. It is known that sensible
ultra-local lattice Dirac operators cannot exactly satisfy the GWR
\cite{nogo}; but approximate solutions are possible, and for these
we have $D_{ov}\approx D$, indicating that $D_{ov}$ approximately
inherits the properties of $D$ and is also likely to have good locality 
properties since $D$ is ultra-local. 

A specific construction of a short range lattice Dirac $D$ with the 
properties (i) and (ii) above arises from the perfect action formalism. 
This formalism produces, in principle, a lattice fermion action free of 
artifacts in the scaling behavior via the iteration of renormalization group 
transformations. Moreover, the corresponding lattice Dirac operator 
satisfies the GWR as well \cite{GW}. The construction of the perfect action 
can only be carried out explicitly for free and for perturbatively 
interacting fermions \cite{QuaGlu}.\footnote{In that case, the axial
anomaly takes the correct form even at finite lattice
spacing \cite{BWPLB}.} Still, at the non-perturbative level one 
can construct {\em approximations} of the {\em classically} perfect action 
(or fixed point action) of asymptotically free theories. 
The scaling artifacts of the fixed point action
tend to be very small --- as a study in 
the 2d $O(3)$ model revealed \cite{o3} --- and the fixed point Dirac
operator solves the GWR too \cite{Has}. Hence approximations, or 
truncations, of the fixed point Dirac operator are natural candidates for
operators with the properties (i) and (ii) above.\footnote{For direct QCD 
applications of a truncated fixed point Dirac operator,
see Ref.\ \cite{Bern,Bern2}. However, the truncation distorts the
chiral symmetry; indeed, even for truncated perfect actions the
additive mass renormalization can be considerable \cite{BBCW}. Chiral 
symmetry can be re-imposed though by inserting the truncated fixed point
operator into the overlap formula, and this is another motivation to
study the overlap operator with truncated fixed point operator as kernel.}
In particular, a {\em hypercubic} approximation of the fixed point Dirac
operator has been considered, and it was found in numerical studies that
the use of this operator as input in the overlap formula can significantly
improve the scaling, locality and convergence properties of the overlap
Dirac operator \cite{Schwing,HFNPB,Bern}.
These properties have been demonstrated in the Schwinger
model \cite{Schwing}, and also in QCD they
have a potential to compensate the additional complication
in $D_{HF}$ compared to $D_{W}$: for the HF described in
Ref.\ \cite{HFNPB} the locality of $D_{ov}$ is improved by
a factor of 2 in the exponential decay 
compared to the standard overlap fermion.
\footnote{This number refers to quenched QCD at $\beta =6$,
and the corresponding test for the standard overlap fermion
was performed in Ref.\ \cite{HJL}.}
Also the convergence rate increases significantly.
However, to establish the overlap HF operator as a viable Dirac operator
for lattice QCD
one also has to check the conceptual basis, in particular whether
the correct axial anomaly and index are reproduced, and this is the issue 
that we address in the present paper.
Experiments with simpler, non-standard operators
inserted in the overlap formula have been performed in Refs.\
\cite{simple}. All those formulations are also covered as special cases
by the considerations in this paper.

The organization of the paper is as follows.
In Section 2 we discuss the properties of $D_{HF}$ which are needed to compute
its axial anomaly and index in Section 3. The conclusions and an outlook on
further generalizations are given in the Section 4.

\section{The structure of the hypercube Dirac operator}

We are going to use the following conventions for the
$\gamma$ matrices: $(\gmu)^{\dagger} = \gmu\,$, 
$\,\{\gmu,\gnu\} = 2\delta_{\mu\nu}\,$, 
$\,\g5 = i^n\,\gamma_1 \cdots \gamma_{2n}\,$, so that
$(\g5)^{\dagger} = \g5$ and 
$\tr (\g5 \gamma_1 \cdots \gamma_{2n}) = (-2i)^{n}$.
Throughout this paper summation over repeated indices is implied.

What we consider here is the minimally gauged HF-Dirac operator, 
which we are going to describe now. For techniques to simulate
such HFs in QCD, see Ref.\ \cite{simu}.

Assume the two lattice sites 
$x$ and $y$ to belong to the same lattice hypercube, i.e.\
$\vert x_{\mu} - y_{\mu} \vert \leq a$ for $\mu = 1 \dots 2n$.
Then we denote by $P(x,y)$ the set of lattice paths of minimal length
connecting $x$ and $y$. All these paths are inside the same hypercube
again, their length is $n_{xy} \in \{ 0,1, \dots , 2n \}$, 
and the number of such paths is $n_{xy}!$.
Let $\sigma$ be one of these paths. In the presence of a compact
lattice gauge field $U_{\mu}(x)$ we denote by
$U(\sigma )$ the product of link variables along the path $\sigma$.
We also define the sign function $\eps (t) ={\rm sign} \, t$ for
$t \neq 0$, and $\eps (0) = 0$. Then the minimally gauged HF 
operator can be written as
\bea
D_{HF} &=& \frac{1}{a} \Big( \gamma_{\mu} \rho_{\mu}
+ \lambda \Big) \ , \label{DHF} \\
(\rho_{\mu})_{xy} &=& \kappa_{n_{xy}} \frac{\eps (x_{\mu}-y_{\mu})}
{n_{xy}!} \sum_{\sigma \in P(x,y)} U(\sigma ) \ , \nonumber \\
\lambda_{xy} &=& \lambda_{n_{xy}} \frac{1}{n_{xy}!}
\sum_{\sigma \in P(x,y)} U(\sigma ) \ . \label{lambda}
\eea
The parameters $\kappa_{1}, \dots \kappa_{2n}$ and $\lambda_{0},
\lambda_{1},\dots , \lambda_{2n}$ are coupling constants.
Here we stay with the simple $\gamma$-structure of the Wilson
fermion. \footnote{HFs with a more general Clifford algebra
have also been used in QCD simulation, see 
Refs.\ \cite{BBCW,Kostas,Bern}.}
The vector term $\rho_{\mu}$ alone characterizes a 
generalized naive fermion, which would by itself
generate species doublers. The scalar term $\lambda $ 
can be considered a generalized Wilson term which removes these
doublers (resp.\ attaches a mass of the cutoff scale to them)
for suitable couplings, see below.
$\kappa_{n_{xy}}$ and $\lambda_{n_{xy}}$ couple one site $x$
to $2^{n_{xy}}$ sites $y$. 
\footnote{The usual Wilson-Dirac
operator with Wilson parameter $r$ and bare mass $m_0$ is recovered 
by setting $\kappa_1\!=\!1/2\,$, $\lambda_0\!=\!2nr\!+\!am_0\,$, 
$\lambda_1\!=\!-r/2$ and $\kappa_j=\lambda_j=0$ for $j\ge2$.}

$D_{HF}$ has the
correct formal continuum limit with vanishing bare mass precisely when
\be \label{constraints}
\sum_{j=1}^{2n} 2^{j}\left[ {\textstyle 
{2n -1 \atop j-1}} \right]
\kappa_{j} = 1 \ , \qquad
\sum_{j=0}^{2n} 2^{j} \left[ {\textstyle 
{2n \atop j}} \right] \lambda_{j} = 0 \ .
\ee
At a finite $\beta$ one deviates from these constraints
and amplifies each coupling (except for $\lambda_{0}$) 
in order to compensate its
suppression by the link variable \cite{HFNPB}. However,
in the current context we do impose the above constraints 
because they have to be restored in the classical continuum limit.
They can be used to eliminate 
$\kappa_1$ and $\lambda_0\,$; then $D_{HF}$ 
contains $4n\!-\!1$ free parameters $\kappa_2,\dots,\kappa_{2n}\,$; 
$\lambda_1,\dots,\lambda_{2n}$.

For our purposes it is useful to express $D_{HF}$ in a coordinate-free way
as follows. We use the parallel transporters
\be
T_{+ \mu}(x,y) = U_{\mu}(x) \delta_{x,y-a\hat \mu} \ , 
\qquad 
T_{- \mu}(x,y) = U_{\mu}^{\dagger}(x- \hat \mu) \delta_{x-a\hat \mu ,y} \ ,
\ee
(where $\hat \mu$ is the unit vector in $\mu$ direction) to
define the Hermitian operators
\be
S_{\mu} = \frac{1}{2i} (T_{+\mu} - T_{-\mu}) \ , \quad
C_{\mu} = \frac{1}{2} (T_{+\mu} + T_{-\mu}) \ .\label{SC}
\ee
Moreover we introduce the following notation for the symmetrized product 
of operators,
\be
[ \O_{1} \dots \O_{p}]_{sym} = \frac{1}{p!} \sum_{\alpha}
\O_{\alpha (1)} \dots \O_{\alpha (p)} \ ,
\ee
where the sum runs over all permutations $\alpha$ of $\{ 1, \dots ,p \}$.
Then 
eqs.\ (\ref{lambda}) can be re-expressed as 
\bea
-i\,\rho_{\mu} &=& S_{\mu}\;-\;\sum_{p=2}^{2n}2^p\kappa_p
\sum_{\nu_2<\cdots<\nu_p\,;\ \nu_j\ne\mu\ \forall{}j}
\lb{}S_{\mu}(1-C_{\nu_2}\cdots{}C_{\nu_p})\rb_{sym} \nn \ , \\
\lambda &=& \sum_{p=1}^{2n}2^p(-\lambda_p)
\sum_{\nu_1<\cdots<\nu_p}\lb{}1-C_{\nu_1}\cdots{}C_{\nu_p}\rb_{sym}
\label{rholam} \ .
\eea
Note that $\rho_{\mu}^{\dagger}=-\rho_{\mu}\,$, $\,\lambda^{\dagger}=\lambda$,
hence $D_{HF}$ satisfies $\gamma_{2n+1}$-Hermiticity,
\be
D_{HF}^{\dagger}=\gamma_{2n+1}\,D_{HF}\,\gamma_{2n+1} \ .
\ee

We now consider the zero-modes of the free field ($U\!=\!1$) ``naive'' 
HF-Dirac operator $\frac{1}{a} \gmu \rho_{\mu}$ and their ``masses'' 
provided by the scalar term 
$\frac{1}{a}\lambda$ in eq.\ (\ref{DHF}).
The free field momentum representation of $S_{\mu}\,$, $C_{\mu}$
(i.e. their eigenvalues for the plane wave eigenfunction $e^{ikx/a}$)
is obvious from eq.\ (\ref{SC}),
\be \label{SCsc}
S_{\mu}(k)=\sin(k_{\mu})\equiv{}s_{\mu} \ , \qquad
C_{\mu}(k)=\cos(k_{\mu})\equiv{}c_{\mu} \ .
\ee 
Hence the free field momentum representations
\footnote{Note that $k$ represents a {\em re-scaled} momentum,
which is $2\pi$ periodic at any lattice spacing $a$.
Nevertheless we denote $k$ simply as ``the momentum''.}
of $\rho_{\mu}$ and $\lambda$ are
\bea
\rho_{\mu}(k)&=&is_{\mu}\Big\lb\,1-\sum_{p=2}^{2n}2^p\kappa_p
\sum_{\nu_2<\cdots<\nu_p\,;\ \nu_j\ne\mu\ \forall{}j}(1-c_{\nu_2}\cdots{}c_{\nu_p})
\Big\rb \label{2.23} \\
\lambda(k)&=&\sum_{p=1}^{2n}2^p(-\lambda_p)
\sum_{\nu_1<\cdots<\nu_p}(1-c_{\nu_1}\cdots{}c_{\nu_p})
\label{2.30}
\eea
The former vanishes when $s_{\mu}\!=\!0$, so the naive HF-Dirac 
operator has the usual 
zero-mode at $k=0$ and the familiar $2^{2n}-1$ ``doubler'' zero-modes 
for $k$ at the
corners of the Brillouin zone, just as in the case of the usual naive 
Dirac operator.
However, in addition to these, there can be other zero-modes 
corresponding to vanishing of the
factor in the square brackets 
on the right-hand side of eq.\ (\ref{2.23}). It can vanish
for some momenta $k$ with components different from $0$ and $\pi$,
when $\kappa_2,\dots,\kappa_{2n}$ are in certain regions 
of the parameter 
space. These zero-modes correspond to new ``exotic'' spurious fermion 
species:
if such a zero-mode occurs at $k=k^{(0)}$ we set $k=k^{(0)}+k'$ and 
find that the 
leading order term in the expansion of $\gmu\rho_{\mu}(k)$ around 
$k^{(0)}$ is
$\sim\,\gmu\sum_{\nu\ne\mu}k_{\nu}'$. The corresponding propagator 
does not describe 
a usual Dirac fermion species. These exotic species are excluded though
if the parameters $\kappa_2,\dots,\kappa_{2n}$ satisfy
\be
\chi_{\mu}\lb\kappa_2,\dots,\kappa_{2n}\rb(k)\,\equiv\,
\sum_{p=2}^{2n}2^p\kappa_p
\sum_{\nu_2<\cdots<\nu_p\,;\ \nu_j\ne\mu\ \forall{}j}
(1-c_{\nu_2}\cdots{}c_{\nu_p})
\;<\;1\qquad\forall\,k \ .
\label{2.24}
\ee
(Note that if this is satisfied for a particular index $\mu$ then it is 
satisfied for all $\mu=1,\dots,2n$. 
Also, since $\chi_{\mu}=0$ at $k=0$, the condition
$\chi_{\mu}<1$ is the same as $\chi_{\mu}\ne1$.)

At this point it is natural to ask what are the values of 
$\kappa_1,\dots,\kappa_{2n}$
that are of interest in practice, and do they satisfy (\ref{2.24})?
The values of $\lambda_1,\dots,\lambda_{2n}$ used in practice are 
also relevant here
since they determine $\lambda(k)$ and hence the masses of the doubler 
fermion species.
As discussed in the introduction, one of the main aims in choosing 
the coupling parameters
is to make $D_{HF}$ as close as possible to satisfying the GW relation. 
The procedure of truncating perfect fermions
(described in Section 1) led to the following values for the 
couplings \cite{BBCW,EPJC} in dimensions 
$2n\!=\!2$ and
$2n\!=\!4\,$:\footnote{We give the values to 3 decimal places; 
they are given to
higher precision in \cite{BBCW,EPJC}.}

\noindent $2n=2\,$:
\bea
\kappa_1=0.309\quad&,&\quad\kappa_2=0.095 \nonumber \\
\lambda_0=1.490\quad,\quad\lambda_1&=&-0.245\quad,\quad\lambda_2=-0.128
\label{2.26}
\eea
$2n=4\,$:
\bea
\kappa_1=0.137\quad,\quad\kappa_2=0.032\quad&,&\quad\kappa_3=0.011
\quad,\quad\kappa_4=0.005 \nonumber \\
\lambda_0=1.853\, ,\ \lambda_1=-0.061\, ,\ \lambda_2&=&-0.030\, ,\
\lambda_3=-0.016\, ,\ \lambda_4=-0.008 \nonumber \\
&&\label{2.27}
\eea
In two dimensions the left-hand side of eq.\ (\ref{2.24}) is
\be
\chi_{\mu}^{(2n=2)}=4\kappa_2(1-c_{\nu}) \, \epsilon_{\mu \nu} \ .
\label{2.28}
\ee
The maximum of this, attained at $c_{\nu}=-1$, is
$8\kappa_2=0.76$ for the coupling values in eq.\ (\ref{2.26}), 
hence eq.\ (\ref{2.24}) is
satisfied. In the dimension four case the left-hand side of eq.\ 
(\ref{2.24}) can be re-written as
\bea
\chi_{\mu}^{(2n=4)}&=&12\kappa_2+24\kappa_3+16\kappa_4
+16\frac{\kappa_3^2}{\kappa_2}\Big(\alpha^3-(\alpha+c_{\nu_2})
(\alpha+c_{\nu_3})
(\alpha+c_{\nu_4})\Big) \nonumber \\
&&\ -16\Big(\kappa_4-\frac{\kappa_3^2}{\kappa_2}\Big)
c_{\nu_2}c_{\nu_3}c_{\nu_4} \ ,
\label{2.29}
\eea
where $\{\mu,\nu_2,\nu_3,\nu_4\}=\{1,2,3,4\}$ 
and $\alpha=\kappa_2/(2\kappa_3)$.
From this we see that when $\kappa_2\ge2\kappa_3$ (i.e. $\alpha\ge1$) and 
$\kappa_4\ge\kappa_3^2/\kappa_2$ the maximum of $\chi_{\mu}^{(2n=4)}$ 
is attained at
$c_{\nu_2}\!=\!c_{\nu_3}\!=\!c_{\nu_4}\!=\!-1$. For the coupling values 
(\ref{2.27}) 
this maximum is $0.93$, hence (\ref{2.24}) is again satisfied.

Let $\{k^{(j)}\}$ denote the momenta of the zero-modes for the free field naive
HF-Dirac operator. The mass of such a mode, provided by the scalar term 
in $D_{HF}\,$,
is $M^{(j)}/a$ where $M^{(j)}\equiv\lambda (k^{(j)})$.
To avoid species doubling in the full HF-Dirac 
operator we impose the requirement on $\lambda_1,\dots,\lambda_{2n}$ that
$\lambda (k^{(j)})>0$ for $k^{(j)}\ne0$. From eq.\ (\ref{2.30}) we see that a 
sufficient condition for this is 
$\lambda_1<0$ and $\lambda_p\le0$ $\ \forall\,p\!=\!2,\dots,2n$, 
which holds for the coupling values in eqs.\ (\ref{2.26})--(\ref{2.27})
(and of course also for the Wilson-Dirac operator).
For the usual zero- and doubler modes, characterized by $s_{\mu}\!=\!0$ 
$\ \forall\,\mu\,$, i.e. $k_{\mu}=0$ or $\pi$ for each $\mu\,$, let
$N_{\pi}$ denote the number of $k_{\mu}$'s equal to $\pi$.
Then, from (\ref{2.30}), the mass $M/a$ of the mode is seen to depend only on
$N_{\pi}\,$ as
\be
M(N_{\pi})=\sum_{p=1}^{2n}2^{p+1}(-\lambda_p)\sum_{q=1}^p
{\textstyle \left\lb{2n-N_{\pi} 
\atop p-q}\right\rb\,\left\lb{N_{\pi} \atop q}\right\rb} \ ,
\label{2.30a}
\ee
with $\left\lb{N_{\pi} \atop q}\right\rb\equiv0$ for $q>N_{\pi}$. 
From this the masses in dimensions 2 and 4, with couplings given by eqs.\
(\ref{2.26})--(\ref{2.27}), can be determined (see also Figures 1 and 2
in Ref.\ \cite{EPJC}). We list them in the Table below in
lattice units. For comparison we also list the masses of 
the corresponding modes
of the Wilson-Dirac operator with Wilson parameter $r$.\\

\noindent $2n=2\,$:

\begin{tabular}{l|c|c|c|}
$N_{\pi}$ & $0$ & $1$ & $2$ \\ \cline{1-4}
$M_{HF}$  & $0$ & $2.004$ & $1.960$ \\ \cline{1-4}
$M_W$     & $0$ & $2r$ & $4r$ \\ 
\end{tabular}

\vspace*{2mm}

\noindent $2n=4\,$:

\begin{tabular}{l|c|c|c|c|c|}
$N_{\pi}$ & $0$ & $1$ & $2$ & $3$ & $4$ \\ \cline{1-6}
$M_{HF}$  & $0$ & $1.988$ & $1.960$ & $1.964$ & $2.000$ \\ \cline{1-6}
$M_W$     & $0$ & $2r$ & $4r$ & $6r$ & $8r$ \\ 
\end{tabular}

\ \\

The HF doubler masses are all close to $2$ 
in lattice units, reflecting 
the fact that the free field $D_{HF}$ with the coupling 
values from eq.\ (\ref{2.26}) resp.\
(\ref{2.27}) are good approximate solutions to the GW relation 
(since for exact GW
solutions the eigenvalues lie on the circle in the complex plane 
centered at $(1/a,0)$ 
with radius $1/a$). We also remark that, unlike in the Wilson case, $M_{HF}$ 
does not always increase with increasing $N_{\pi}$.


\section{The continuum limit of the axial anomaly for
the overlap-HF Dirac operator}

For a given value of the parameter $m$, the momenta of the 
zero-modes of the free field $D_{ov}$
are the $k^{(j)}$ with $M^{(j)}<m$ (both defined in Section 2).
Hence the parameter
region in which $D_{ov}$ has a physical zero-mode and no doublers is
$0<m<\mbox{min}\{M^{(j)}\ne0\}$. 

Our aim now is to evaluate the classical continuum limit of the 
topological charge
density $q(x)$ given by eq.\ (\ref{density}), or equivalently, the axial 
anomaly $\A(x)=2iq(x)$,
of the overlap-HF Dirac operator. Specifically, we consider the situation where
$D_{ov}$ is coupled to the lattice transcript of a smooth continuum gauge field
$A=A_{\mu}(x)dx^{\mu}$, i.e. the link variables are 
\be
U_{\mu}(x)=T\exp\Big(a \int_0^1 A_{\mu}(x+(1-t)a\hat{\mu})\,dt\Big) \ ,
\label{3.2}
\ee
where $T$ implies $t$-ordering. We will derive the following result:\\

{\em If $m\ne{}M^{(j)}\,$ $\ \forall\,j$\, ,
then
\be
\lim_{a\to0}\ q(x)=I(\kappa_2,\dots,\kappa_{2n};
\lambda_1,\dots,\lambda_{2n};m)
\,q_{cont}(x)
\label{3.3}
\ee
where
\be
q_{cont}(x)=\frac{1}{(2\pi{}i)^nn!}\,\cdot\,\frac{1}{2^n}\,
\e_{\mu_1\dots\mu_{2n}}
\,\tr \Big[ F_{\mu_1\mu_2}(x)\cdots{}F_{\mu_{2n-1}\mu_{2n}}(x) \Big]
\label{3.4}
\ee
is the continuum topological charge density and $I(\kappa_2,
\dots,\lambda_{2n};m)$
is the degree of a certain map $\Theta:T^{2n}\to{}S^{2n}$ given in eq.\
(\ref{3.26}) below. In particular, $I(\kappa_2,\dots,\lambda_{2n};m)=1$ 
holds for $m$ in 
the physical (doubler-free) region $0<m<\mbox{min}\{M^{(j)}\ne0\}$. 
Thus, for $m$
in this region, $q(x)$ and the axial anomaly reduce to the correct continuum 
expressions in the classical continuum limit. Furthermore, when the parameters
$\kappa_2,\dots,\kappa_{2n}$ satisfy the constraint (\ref{2.24}), then}
\be 
I(\kappa_2,\dots,\kappa_{2n};\lambda_1,\dots,\lambda_{2n};m)
=\sum_{\{N_{\pi}\,:\,M(N_{\pi})<m\}}{\textstyle \left\lb{2n \atop 
N_{\pi}}\right\rb} \,(-1)^{N_{\pi}} \ .
\label{3.5}
\ee

We first derive the result in the infinite volume, 
i.e.\ on a hypercubic lattice on ${\R}^{2n}$, and thereafter  
we discuss the finite volume $2n$-torus case. 
The expression (\ref{density}) can be re-written as 
\be
q(x)=-\frac{1}{2}\,\tr\Big(\frac{H_m}{\sqrt{H_m^2}}\Big)(x,x) \ ,
\label{3.7}
\ee
where 
\be
H_m=\g5(aD_{HF}-m)=\g5(\gmu\rho_{\mu} + \lambda - m)
\label{3.6}
\ee
is the Hermitian HF-Dirac operator (normalized by $1/a$). 
We proceed as in the Wilson case treatment of Refs.\ \cite{DA(AP),DA(JMP)} 
by expanding
$(H_m^2)^{-1/2}$ as a power series. First, $H_m^2$ is decomposed as
\bea
H_m^2 &=& L-V \nn \\
L=-\rho^2+(\lambda-m)^2\ &, &\ V=\gmu\lb\rho_{\mu}\,,\lambda\rb
+{\textstyle \frac{1}{2}}\gmu\gnu\lb\rho_{\mu}\,,\rho_{\nu}\rb \ .
\label{3.9}
\eea
As in the Wilson case 
we observe $\Vert V \Vert \sim{} O(a^2)$, 
which is a consequence of the property 
$\Vert \, \lb{}\,T_{\pm\mu}\,,T_{\pm\nu}\rb\, \Vert \sim{}O(a^2)$ and 
eqs.\ (\ref{rholam}). Furthermore, a lower bound
$0<b<L$ exists when the lattice is sufficiently fine.
\footnote{This was established
in the Wilson case for restricted values of $m$ in Ref.\ \cite{HJL,
Neuloc}, and 
later for general $m$ in \cite{DA(bound)}. The result will be 
generalized to the 
present HF case, and more general cases, in \cite{prep}.} 
This implies that
$||L^{-1}V||\sim{}O(a^2)$ and consequently 
$(H_m^2)^{-1/2}=(L[1-L^{-1}V])^{-1/2}$ 
can be expanded as a power series in $L^{-1}V$ 
when the lattice spacing $a$ is sufficiently small.
This was done in the Wilson case \cite{DA(AP),DA(JMP)}
using an integral representation of the inverse square root.
(Note that the integral representation is needed since
$L$ and $V$ do in general not commute.)
The argument relies only on general properties of $L$ 
and $V$ which continue to hold in the present HF case;
the treatment in Refs.\ \cite{DA(AP),DA(JMP)} generalizes 
straightforwardly to the HF and to arbitrary
even dimension.
Substituting the resulting expansion into 
eq.\ (\ref{3.7}) and using the lattice $\delta$-function
\be
\delta_x=\int_{-\pi}^{\pi}\frac{d^{2n}k}{(2\pi)^{2n}}\,e^{-ikx/a}\,\phi_k
\ , \qquad \phi_k(y)\equiv{}e^{iky/a}\ ,
\label{3.10}
\ee
to express $q(x)$ as an integral over momentum space, one obtains
\footnote{To derive eq.\
(\ref{3.11}) we have used the fact that the trace of the product of $\g5$ with 
the product of less than $2n$ $\gamma$ matrices vanishes. 
The factor $1/a^{2n}$ in eq.\
(\ref{3.11}) originates from the $a^{2n}$ in the operator representation
$\O\psi(x)=a^{2n}\sum_y\O(x,y)\psi(y)\,$ ($\O=H_m/\sqrt{H_m^2}\ $).
Hence the first term is of $O(1)$.}
\be
q(x)=-\frac{1}{2}\,c(n)\,\frac{1}{a^{2n}}\int_{-\pi}^{\pi}
\frac{d^{2n}k}{(2\pi)^{2n}}\,
\frac{\tr(H_m(k)\,e^{-ikx/a}\,V^n\,e^{ikx/a})}{L(k)^{n+1/2}}+O(a) \ ,
\label{3.11}
\ee
where $H_m(k)$ and $L(k)$ are the free field momentum representations 
of $H_m$ and $L$, and
\be
c(n)=\frac{1}{n!}\,\frac{d^{n}}{dt^{n}}\Big(1-t\Big)^{-1/2}\Big|_{t=0}
=\frac{(2n)!}{2^{2n}(n!)^2} \ .
\label{3.12}
\ee

To evaluate the limit $a\to0$ of eq.\ (\ref{3.11}) 
we start from the following general observations,
\bea
e^{-ikx/a}\,\lb{}T_{\pm\mu}\,,T_{\pm\nu}\rb\,e^{ikx/a}
&=&a^2F_{\mu\nu}(x)\,e^{i(\pm{}k_{\mu}\pm{}k_{\nu})}+O(a^3) \ , \nn \\
e^{-ikx/a}\,\lb{}T_{\pm\mu}\,,T_{\mp\nu}\rb\,e^{ikx/a}
&=&-a^2F_{\mu\nu}(x)\,e^{i(\pm{}k_{\mu}\mp{}k_{\nu})}+O(a^3) \ ,
\label{3.14}
\eea
which imply
\bea
e^{-ikx/a}\,\lb{}S_{\mu}\,,S_{\nu}\rb\,e^{ikx/a}
&=&-a^2F_{\mu\nu}(x)\,c_{\mu}c_{\nu}+O(a^3) \ , \nn \\
e^{-ikx/a}\,\lb{}S_{\mu}\,,C_{\nu}\rb\,e^{ikx/a}
&=&a^2F_{\mu\nu}(x)\,c_{\mu}s_{\nu}+O(a^3) \ , \nn \\
e^{-ikx/a}\,\lb{}C_{\mu}\,,C_{\nu}\rb\,e^{ikx/a}
&=&-a^2F_{\mu\nu}(x)\,s_{\mu}s_{\nu}+O(a^3) \ , \label{3.17}
\eea
with the terms defined in eq.\ (\ref{SCsc}).
In the following we denote the free field momentum representation of a general
lattice operator $X$ by $X(k)$. Then the relations (\ref{3.17})
can be expressed collectively as 
\bea
e^{-ikx/a}\,\lb{}X\,,Y\rb\,e^{ikx/a}
&=&-a^2F_{\alpha\beta}(x)\,\partial_{\alpha}X(k)\,\partial_{\beta}Y(k)+O(a^3) 
\label{3.18}
\eea
for $X=S_{\mu}\,,C_{\mu}$ and $Y=S_{\nu}\,,C_{\nu}$. 
In fact this relation continues 
to hold when $X$ and $Y$ are general polynomials of the 
$S_{\mu}$ and the $C_{\mu}$.
Since $\rho_{\mu}$ and $\lambda$ are such polynomials 
we can apply eq.\ (\ref{3.18}) to 
$e^{-ikx/a}\,V^n\,e^{ikx/a}$ in the expression (\ref{3.11}) to obtain
\bea
&& \hspace*{-15mm} \tr(H_m(k)\,e^{-ikx/a}\,V^n\,e^{ikx/a}) \nonumber \\
&& \hspace*{-15mm} = 
i^n\,a^{2n}\,\e_{\mu_1\dots\mu_{2n}}\,\tr\Big(F_{\alpha_1\alpha_2}(x)
\cdots{} F_{\alpha_{2n-1}\alpha_{2n}}(x)\Big) \times \nn \\
&& \hspace*{-15mm} \Big\lb\,(\lambda(k)-m)\,\partial_{\alpha_1}
\rho_{\mu_1}(k)\cdots\partial_{\alpha_{2n}}\rho_{\mu_{2n}}(k) \nonumber \\
&&\hspace*{-15mm} \quad 
-2n\,\partial_{\alpha_1}\rho_{\mu_1}(k)\cdots\partial_{\alpha_{2n-1}}
\rho_{\mu_{2n-1}}(k)\,\partial_{\alpha_{2n}}\lambda(k)\,\rho_{\mu_{2n}}(k)\Big\rb
+O(a^{2n+1}) \ .
\label{3.19}
\eea
We now note the two identities, which will be crucial for our further
considerations:
\bea
&& \hspace*{-17mm}
\e_{\mu_1\dots\mu_{2n}}\,\tr\Big(F_{\alpha_1\alpha_2}(x)\cdots{}
F_{\alpha_{2n-1}\alpha_{2n}}(x)\Big)\,\partial_{\alpha_1}
\rho_{\mu_1}(k)\cdots\partial_{\alpha_{2n}}\rho_{\mu_{2n}}(k) = \nonumber \\
&& \hspace*{-17mm} \e_{\mu_1\dots\mu_{2n}}\,\tr\Big(F_{\mu_1\mu_2}(x)\cdots{}
F_{\mu_{2n-1}\mu_{2n}}(x)\Big)\,\e_{\alpha_1,\dots\alpha_{2n}}\,
\partial_{\alpha_1}\rho_1(k)\cdots\partial_{\alpha_{2n}}\rho_{2n}(k) \ ,
\label{3.20}
\eea
\bea
&& \hspace*{-15mm} 
\e_{\mu_1\dots\mu_{2n}}\,\tr\Big(F_{\alpha_1\alpha_2}(x)\cdots{}
F_{\alpha_{2n-1}\alpha_{2n}}(x)\Big) \times \nonumber \\
&& \hspace*{-15mm}
\quad\ \partial_{\alpha_1}\rho_{\mu_1}(k)\cdots\partial_{\alpha_{2n-1}}
\rho_{\mu_{2n-1}}(k)\,\partial_{\alpha_{2n}}\lambda(k)\,\rho_{\mu_{2n}}(k) =
\nonumber \\
&& \hspace*{-15mm}
\e_{\mu_1\dots\mu_{2n}}\,\tr\Big(F_{\mu_1\mu_2}(x)\cdots{}
F_{\mu_{2n-1}\mu_{2n}}(x)\Big)\,
\sum_{p=1}^{2n}(-1)^p\,\rho_p(k)\,\e_{\alpha_0\alpha_1\dots
\alpha_{p-1}\alpha_{p+1}\dots\alpha_{2n}} \times \nonumber \\
&& \hspace*{-15mm} \quad
\partial_{\alpha_0}\lambda(k)\,\partial_{\alpha_1}\rho_1(k)\cdots
\partial_{\alpha_{p-1}}\rho_{p-1}(k)\,\partial_{\alpha_{p+1}}\rho_{p+1}(k)
\cdots\partial_{\alpha_{2n}}\rho_{2n}(k) 
\label{3.21}
\eea
These combinatorial identities rely only on the facts that 
the $\rho_{\mu}(k)$ and $\lambda (k)$ all commute, on
$F_{\mu\nu}(x)\!=\!-F_{\nu\mu}(x)$, and on the cyclic 
property of the trace. 
Replacing the
left-hand sides of these identities by the right-hand sides in eq.\
(\ref{3.19}), and introducing the real-valued functions
\be
\theta_0(k)=-[\lambda (k)-m]\quad,\qquad\theta_{\mu}(k)=-i\,\rho_{\mu}(k)
\qquad\ \mbox{($\mu\!=\!1,\dots,2n$)}\,,
\label{3.22}
\ee
we arrive at
\bea
&& \hspace*{-10mm} \tr \Big[ H_m(k)\,e^{-ikx/a}\,V^n\,e^{ikx/a} \Big]
\,dk_1\wedge\dots\wedge{}dk_{2n} = \nonumber \\
&& \hspace*{-15mm} 
-(-i)^n\,a^{2n}\,\e_{\mu_1\dots\mu_{2n}}\,\tr\Big(F_{\mu_1\mu_2}(x)\cdots{}
F_{\mu_{2n-1}\mu_{2n}}(x)\Big) \times \nonumber \\
&& \hspace*{-15mm} \quad 
\sum_{p=0}^{2n}(-1)^p\,\theta_p\,d\theta_0\wedge\dots\wedge{}
d\theta_{p-1}\wedge{}d\theta_{p+1}\wedge\dots\wedge{}d\theta_{2n}
+ O(a^{2n+1}) \ ,
\label{3.23}
\eea
where $d\theta_j\equiv\partial_{\alpha}\theta_j\,dk_{\alpha}$ 
is the {\em exterior derivative} of $\theta_j$. 
Substituting this into eq.\ (\ref{3.11}) and re-writing $L$ 
from eq.\ (\ref{3.9}) as
\be
L(k)=-\rho^2(k)+ [ \lambda (k)-m ]^2
= \sum_{p=0}^{2n}\theta_p(k)^2=|\theta(k)|^2 \ ,
\ee
we obtain $\lim_{a\to0}\ q(x)=I(\kappa_2,\dots,\lambda_{2n};m)\,q_{cont}(x)$,
as we claimed in eq.\ (\ref{3.3}), with
\bea
&& \hspace*{-20mm} I(\kappa_2,\dots,\lambda_{2n};m) = 
\frac{1}{2}\,c(n)\,\frac{n!}{\pi^n}
\times \nn \\
&& \hspace*{-17mm} \int_{\;\lb-\pi\,,\,\pi\rb^{2n}}\frac{1}{|\theta|^{2n+1}}
\sum_{p=0}^{2n}(-1)^p\,\theta_p\,d\theta_0\wedge\dots\wedge{}
d\theta_{p-1}\wedge{}d\theta_{p+1}\wedge\dots\wedge{}d\theta_{2n} \ .
\label{3.24}
\eea
A little calculation shows that the integrand here can be re-written as 
\be
\sum_{p=0}^{2n}(-1)^p\,\frac{\theta_p}{|\theta|}\,
d\Big(\frac{\theta_0}{|\theta|}\Big)
\wedge\dots\wedge{}d\Big(\frac{\theta_{p-1}}{|\theta|}\Big)
\wedge{}d\Big(\frac{\theta_{p+1}}{|\theta|}\Big)
\wedge\dots\wedge{}d\Big(\frac{\theta_{2n}}{|\theta|}\Big) \ .
\label{3.25}
\ee
This is precisely the pull-back to $T^{2n}=\rb-\pi\,,\,\pi\rb^{2n}$ 
of the volume 
form on the unit $2n$-sphere $S^{2n}\subset \R^{2n+1}$ via the map
\be
\Theta:T^{2n}\,\to\,S^{2n}\quad,\qquad
\Theta(k):=\Big(\,\frac{\theta_0}{|\theta|}\,,
\frac{\theta_1}{|\theta|}\,,\dots,\,
\frac{\theta_{2n}}{|\theta|}\Big) \ .
\label{3.26}
\ee
Furthermore, the coefficient of the integral in eq.\ (\ref{3.24}) 
turns out to be (recall definition (\ref{3.12}))
\be
\frac{1}{2}\,c(n)\,\frac{n!}{\pi^n}=\frac{(2n)!}{2^{2n+1}\,n!\,\pi^n}
=\frac{1}{Vol(S^{2n})} \ ,
\label{3.27}
\ee
where $Vol(S^{2n})$ is the volume of the unit $2n$-sphere.
Hence expression (\ref{3.24}) calculates the {\em degree} of the map 
(\ref{3.26}),
\be
I(\kappa_2,\dots,\lambda_{2n};m)=deg(\Theta) \ .
\label{3.28}
\ee
This is a generalization of the topological evaluation of the
anomaly coefficient given in Ref.\ \cite{FNS}.

If $\frac{\theta}{|\theta|}\in{}S^{2n} \subset \R^{2n+1}$ 
is a regular point for the
map $\Theta$ then it is a standard topological fact that 
$deg(\Theta)=\sum_ls_l$ , where $s_l=\pm1$ is the sign (relative to 
$dk_1\wedge\dots\wedge{}dk_{2n}$) of the integrand of eq.\ (\ref{3.24}) 
evaluated at a
pre-image $k^{(l)}$ of $\frac{\theta}{|\theta|}\,$, and the sum is over 
all the pre-images (labelled by $l$). 
We choose $\frac{\theta}{|\theta|}=(1,0,\dots,0)$.
Then the pre-images
$k^{(l)}$ are precisely the subset of the $k^{(j)}$ introduced in Section 2
which satisfy 
$\theta_0\!=\!-[\lambda (k)-m] > 0$, i.e.\ for which $M^{(j)}\!
=\!\lambda (k^{(j)})<m$.
Moreover, the integrand in eq.\ (\ref{3.24}) reduces at these momenta to
$d\theta_1\wedge\dots\wedge{}d\theta_{2n}$. To determine the sign of 
this at a  given $k^{(j)}$, recall from eq.\ (\ref{2.23}) that 
$\theta_{\mu}(k)=-i\rho_{\mu}(k)=S_{\mu}(k)[ 1-\chi_{\mu}(k)] $ with 
$\chi_{\mu}(k)$ as defined in eq.\ (\ref{2.24}). It follows that
\bea
d\theta_1\wedge\dots\wedge{}d\theta_{2n}&=&\Big(\,\prod_{\mu=1}^{2n}
C_{\mu}(k)\Big)
\Big(\,\prod_{\nu=1}^{2n}(1-\chi_{\nu}(k))\Big)\,
dk_1\wedge\dots\wedge{}dk_{2n}
\nonumber \\
&&\ +\ \mbox{terms with at least one $S_{\mu}(k)$ factor.}
\label{3.29}
\eea
If $0<m<\mbox{min}\{M^{(j)}\ne0\}$ then there is precisely 
one $k^{(j)}$ for which
$M^{(j)}<m$, namely $k^{(j)}\!=\!0$. In this case, since 
$\chi_{\mu}(0)\!=\!0$ $\ \forall\,\mu\,$, eq.\ (\ref{3.29}) yields
$d\theta_1\wedge\dots\wedge{}d\theta_{2n}\Big|_{k=0}=
dk_1\wedge\dots\wedge{}dk_{2n}\Big|_{k=0}\,$, i.e.\ the sign is $+1$, so 
$deg(\Theta)=1$ in this case, as we claimed. 
Let us now consider the case where 
the parameters $\kappa_2,\dots,\kappa_{2n}$ satisfy the constraint 
(\ref{2.24}), i.e.\ 
$\chi_{\mu}(k)<1$ $\ \forall\,\mu\,,k$. Then the $k^{(j)}$ are precisely
the $k$ at which $S_{\mu}(k)\!=\!0$ $\ \forall\,\mu\,$, so the terms with 
$S_{\mu}(k)$ factors in eq.\
(\ref{3.29}) vanish and the sign of the remainder is given by
$\prod_{\mu=1}^{2n}C_{\mu}(k^{(j)})$. This sign is $(-1)^{N_{\pi}^{(j)}}$ 
where $N_{\pi}^{(j)}$ is the number of components of
$k^{(j)}$ which are equal to $\pi$.
Recalling from Section 2 that $M^{(j)}$ depends only 
on $N_{\pi}^{(j)}$ in this case,
and noting that the number of $k^{(j)}$ with $N_{\pi}^{(j)}=N_{\pi}$ is
$\left\lb{2n \atop N_{\pi}}\right\rb\,$, it follows that
$deg(\Theta)=\sum_{\{N_{\pi}\,:\,M(N_{\pi})<m\}}\left\lb{2n \atop N_{\pi}}\right\rb
(-1)^{N_{\pi}}$. This completes the derivation of the claimed result 
(\ref{3.3})--(\ref{3.5}) in the infinite lattice setting.

The rigorous derivation of the expansion (\ref{3.11}), carried out 
along the same lines as in the overlap-Wilson case \cite{DA(AP),DA(JMP)}, 
requires an assumption on the continuum
field, namely that $A_{\mu}(x)$ and its first few partial derivatives 
are bounded on $\R^{2n}$. 
Such bounds are guaranteed to exist if $A$ has a compact support
on $\R^{2n}$. However, having established the result in eqs.\
(\ref{3.3})--(\ref{3.5}) for
gauge fields with compact support, it can then be extended to 
general smooth gauge fields
using locality type arguments, in the same way as in the overlap-Wilson case
(see eq.\ (3.45) in Ref.\ \cite{DA(AP)} 
and the associated discussion). This relies on the
existence of a non-zero lower bound on $H_m^2\,$. 

In the finite volume $2n$-torus setting the momentum integrals in eqs.\ 
(\ref{3.10}), (\ref{3.11}) become sums, so the derivation given above
does not carry over directly to that setting. However, using 
locality-based arguments
one can show that the finite volume $q(x)$ coincides with the 
infinite volume $q(x)$
up to exponentially suppressed finite size effects,
thereby establishing 
that the continuum limit
results (\ref{3.3})--(\ref{3.5}) continue to hold in the finite 
volume setting.
This was done in the overlap-Wilson case in Ref.\ \cite{DA(JMP)}; the 
arguments there
relied only on general properties and carry over to the present HF case
(given the aforementioned lower bound on $H_m^2$); the details of 
all this will be given 
in the general setting in Ref.\ \cite{prep}. 

In the light of the index formula (\ref{index}) it then follows that
when the overlap-HF operator is coupled to the lattice transcript of a smooth 
continuum gauge field $A$ on $T^{2n}$ with topological charge $Q$, then
$index(D_{ov})$ reduces to $I(\kappa_2,\dots,\lambda_{2n};m)\,Q$ 
in the limit $a\to0$.
Thus the HF fermionic topological charge reduces to the continuum 
topological charge
in the classical continuum limit when the parameter $m$ is in 
the doubler-free region, just as it does in the Wilson case.

\section{Summary}

We repeat that the previous literature contains the following
considerations about the axial anomaly of overlap fermions:
\begin{itemize}

\item Perturbative considerations show that the correct continuum
limit is obtained in the sector of topological charge zero for any
overlap operator, see in particular Refs.\ \cite{RR}.

\item There was also a rigorous, non-perturbative proof that covers 
all topological sectors, but it was specific to the case of the
simplest standard overlap fermion, which uses the Wilson-Dirac 
operator as an input \cite{DA(AP),DA(JMP)}.

\end{itemize}

The standard overlap operator is wide-spread in recent
simulations. However, there are attempts by various groups to
use also non-standard overlap operators 
\cite{Schwing,HFNPB,Bern,simple}. The operators used in those
works are all included in the class of HF overlap operators. For the
latter we have given in this paper a non-perturbative evaluation of
the continuum limit of the axial anomaly and index which is valid in 
all topological sectors.

We have formulated the HF-Dirac operator
in $2n$-dimensional
Euclidean space in the form (\ref{rholam}),
which is well-suited to analytic investigations. 
We used it first to study the dependence of the doubler structure of $D_{HF}$ 
on its coupling parameters. 
Then we evaluated the classical continuum 
limits of the axial anomaly and index of the
overlap-HF Dirac operator, showing that the correct continuum expressions 
are recovered
when parameters are in the physical (doubler-free) region.
A noteworthy feature of our continuum limit evaluation is that it 
relies only 
on general properties of the HF-Dirac operator and not its explicit form. 
This is 
in contrast to the previous evaluations in the Wilson case (a 
special case of the  more general HF structure) which all use the explicit
form of the operator. The main new technical observations which our 
approach is based on are the general relation (\ref{3.18}) and the identities 
(\ref{3.20})--(\ref{3.21}). These ingredients allow the continuum form 
$\e_{\mu_1\dots\mu_{2n}}\,\tr\,F_{\mu_1\mu_2}(x)\cdots{}
F_{\mu_{2n-1}\mu_{2n}}(x)$
of the axial anomaly to be extracted, and its coefficient to be 
topologically evaluated
as the degree of a map $\Theta:T^{2n}\to{}S^{2n}$ using only
general properties
of the HF-Dirac operator. These properties are not specific to the HF 
case, and 
the approach can be extended to completely general overlap Dirac operators
obtained
by substituting a general ultra-local lattice Dirac operator (involving 
the full
Clifford algebra of $\gamma$ matrices) into the overlap formula 
(\ref{over}), as it was done in Ref.\ \cite{Bern}.
The full extension, which involves considerable additional work, 
will be carried out in a forthcoming paper \cite{prep}, 
and it will be shown there that the axial
anomaly and index of the general overlap Dirac operator $D_{ov}$ 
continue to have
the correct classical continuum limits in the physical (doubler-free) 
parameter region specified by $0<m<\mbox{min}\{M^{(j)}\ne0\}$. 
It was worthwhile to consider the overlap-HF case on its own, 
firstly because of the current interest in using this
operator in numerical simulations, and secondly because it illustrates the
main ideas and techniques of the general case but without the extensive
formalism and additional complications of the latter.\\


{\bf\small Acknowledgements} {\small D.A. thanks the lattice group 
at Humboldt Universit\"{a}t,
and W.B. thanks the Lorentz Instituut at Leiden University., 
for hospitality during visits while this work was in progress. 
D.A. is supported by a Marie Curie fellowship 
from the European Commission (contract HPMF-CT-2002-01716), and
W.B. is supported in part by the DFG Sonderforschungsbereich
Transregio 9, ``Computergest\"{u}tzte Theoretische 
Teilchenphysik''.}


\end{document}